\documentclass{PoS}

\usepackage{amsmath}
\usepackage{url}

\usepackage{ifthen} % for conditional statements
\newboolean{uprightparticles}
\setboolean{uprightparticles}{true} %Set to false to get italic particle symbols
\usepackage{xspace}
\usepackage{amsfonts}
\usepackage{upgreek}

\newboolean{pdflatex}
\setboolean{pdflatex}{true} % use this if using eps figures

\def\lhcb {LHCb\xspace}
\def\ux85 {UX85\xspace}
\def\cern {CERN\xspace}
\def\lhc {LHC\xspace}

\def\tt     {TT\xspace}

\ifthenelse{\boolean{uprightparticles}}%
{

 \def\Ppi         {\ensuremath{\uppi}\xspace}

 \def\PDelta      {\ensuremath{\Delta}\xspace}                 
 \def\PXi      {\ensuremath{\Xi}\xspace}                 
 \def\PLambda      {\ensuremath{\Lambda}\xspace}                 
 \def\PSigma      {\ensuremath{\Sigma}\xspace}                 
 \def\POmega      {\ensuremath{\Omega}\xspace}                 
 \def\PUpsilon      {\ensuremath{\Upsilon}\xspace}                 
 
 \def\PB      {\ensuremath{\mathrm{B}}\xspace}                 
                  
 \def\PD      {\ensuremath{\mathrm{D}}\xspace}

 \def\PK      {\ensuremath{\mathrm{K}}\xspace}

 \def\Ph      {\ensuremath{\mathrm{h}}\xspace}                 
 \def\Pi      {\ensuremath{\mathrm{i}}\xspace}

 \def\Pp      {\ensuremath{\mathrm{p}}\xspace}

}
{

 \def\Ppi         {\ensuremath{\pi}\xspace}

 \mathchardef\PDelta="7101
 \mathchardef\PXi="7104
 \mathchardef\PLambda="7103
 \mathchardef\PSigma="7106
 \mathchardef\POmega="710A
 \mathchardef\PUpsilon="7107
                  
 \def\PB      {\ensuremath{B}\xspace}                 
                  
 \def\PD      {\ensuremath{D}\xspace}

 \def\PK      {\ensuremath{K}\xspace}

 \def\Ph      {\ensuremath{h}\xspace}                 
 \def\Pi      {\ensuremath{i}\xspace}

 \def\Pp      {\ensuremath{p}\xspace}

}

\def\c     {\ensuremath{\Pc}\xspace}
\def\cbar  {\ensuremath{\overline \c}\xspace}
\def\ccbar {\ensuremath{\c\cbar}\xspace}

\def\pion  {\ensuremath{\Ppi}\xspace}

\def\pip   {\ensuremath{\pion^+}\xspace}
\def\pim   {\ensuremath{\pion^-}\xspace}
\def\pipi  {\ensuremath{\pion^+\pion^-}\xspace}

\def\kaon  {\ensuremath{\PK}\xspace}
  \def\Kbar  {\kern 0.2em\overline{\kern -0.2em \PK}{}\xspace}

\def\Kz    {\ensuremath{\kaon^0}\xspace}
\def\Kzb   {\ensuremath{\Kbar^0}\xspace}
\def\KzKzb {\ensuremath{\Kz \kern -0.16em \Kzb}\xspace}
\def\Kp    {\ensuremath{\kaon^+}\xspace}
\def\Km    {\ensuremath{\kaon^-}\xspace}

\def\KpKm  {\ensuremath{\Kp \kern -0.16em \Km}\xspace}

  \def\Dbar    {\kern 0.2em\overline{\kern -0.2em \PD}{}\xspace}
\def\D       {\ensuremath{\PD}\xspace}

\def\Dz      {\ensuremath{\D^0}\xspace}
\def\Dzb     {\ensuremath{\Dbar^0}\xspace}
\def\DzDzb   {\ensuremath{\Dz {\kern -0.16em \Dzb}}\xspace}
\def\Dp      {\ensuremath{\D^+}\xspace}
\def\Dm      {\ensuremath{\D^-}\xspace}

\def\DpDm    {\ensuremath{\Dp {\kern -0.16em \Dm}}\xspace}

\def\Dstarp  {\ensuremath{\D^{*+}}\xspace}

\def\B       {\ensuremath{\PB}\xspace}
  \def\Bbar    {\kern 0.18em\overline{\kern -0.18em \PB}{}\xspace}

\def\Bz      {\ensuremath{\B^0}\xspace}

\def\Bs      {\ensuremath{\B^0_s}\xspace}

  \def\Y#1S{\ensuremath{\PUpsilon{(#1S)}}\xspace}% no space before {...}!

\def\proton      {\ensuremath{\Pp}\xspace}

\newcommand{\decay}[2]{\ensuremath{#1\!\to #2}\xspace}         % {\Pa}{\Pb \Pc}

\def\to                 {\ensuremath{\rightarrow}\xspace}

\def\CP                {\ensuremath{C\!P}\xspace}

\def\AT#1     {\ensuremath{A_T^{#1}}\xspace}           % 2

\def\C#1      {\ensuremath{\mathcal{C}_{#1}}\xspace}                       % 9
\def\Cp#1     {\ensuremath{\mathcal{C}_{#1}^{'}}\xspace}                    % 7
\def\Ceff#1   {\ensuremath{\mathcal{C}_{#1}^{\mathrm{(eff)}}}\xspace}        % 9  
\def\Cpeff#1  {\ensuremath{\mathcal{C}_{#1}^{'\mathrm{(eff)}}}\xspace}       % 7
\def\Ope#1    {\ensuremath{\mathcal{O}_{#1}}\xspace}                       % 2
\def\Opep#1   {\ensuremath{\mathcal{O}_{#1}^{'}}\xspace}                    % 7

\def\agamma     {\ensuremath{A_{\Gamma}}\xspace}

\def\kk         {\ensuremath{\PK\PK}\xspace}

\newcommand{\ket}[1]{\ensuremath{|#1\rangle}}              % {b}
\newcommand{\tev}{\ensuremath{\mathrm{\,Te\kern -0.1em V}}\xspace}
\newcommand{\gev}{\ensuremath{\mathrm{\,Ge\kern -0.1em V}}\xspace}
\newcommand{\mev}{\ensuremath{\mathrm{\,Me\kern -0.1em V}}\xspace}
\newcommand{\kev}{\ensuremath{\mathrm{\,ke\kern -0.1em V}}\xspace}
\newcommand{\ev}{\ensuremath{\mathrm{\,e\kern -0.1em V}}\xspace}
\newcommand{\gevc}{\ensuremath{{\mathrm{\,Ge\kern -0.1em V\!/}c}}\xspace}
\newcommand{\mevc}{\ensuremath{{\mathrm{\,Me\kern -0.1em V\!/}c}}\xspace}
\newcommand{\gevcc}{\ensuremath{{\mathrm{\,Ge\kern -0.1em V\!/}c^2}}\xspace}
\newcommand{\gevgevcccc}{\ensuremath{{\mathrm{\,Ge\kern -0.1em V^2\!/}c^4}}\xspace}
\newcommand{\mevcc}{\ensuremath{{\mathrm{\,Me\kern -0.1em V\!/}c^2}}\xspace}

\def\invfb   {\ensuremath{\mbox{\,fb}^{-1}}\xspace}

\def\fs   {\ensuremath{\rm \,fs}\xspace}

\newcommand{\chisq}{\ensuremath{\chi^2}\xspace}

\def\gsim{{~\raise.15em\hbox{$>$}\kern-.85em
          \lower.35em\hbox{$\sim$}~}\xspace}
\def\lsim{{~\raise.15em\hbox{$<$}\kern-.85em
          \lower.35em\hbox{$\sim$}~}\xspace}

\def\sqs   {\ensuremath{\protect\sqrt{s}}\xspace}

\def\pt         {\mbox{$p_T$}\xspace}

\def\tell1  {TELL1\xspace}
\def\ukl1   {UKL1\xspace}

\newcommand{\ie}{\mbox{\itshape i.e.}\xspace}

\def\deltam     {\ensuremath{\Delta m}\xspace}

\def\hphm       {\ensuremath{\Ph^+ \Ph^-}\xspace}

\def\gammahat   {\ensuremath{\hat{\Gamma}}\xspace}

\def\agammadefnot  {\ensuremath{\frac{\gammahat(\Dz \to f) - \gammahat(\Dzb \to f)}{\gammahat(\Dz \to f) + \gammahat(\Dzb \to f)}}\xspace}
\def\etacp      {\ensuremath{\eta_{\CP}}\space}
\def\agammaexp  {\ensuremath{\etacp\left[\frac{1}{2}(A_m+A_d)y\cos\phi-x\sin\phi\right]}\xspace}

\newcommand{\optionalBar}[1]{\ensuremath{\overset{_{(-)}}{#1}}\xspace}

\def\dzkk       {\decay{\Dz}{\Kp\Km}}
\def\dzpipi     {\decay{\Dz}{\pip\pim}}

\def\dstdzpi    {\decay{\Dstarp}{\Dz\pip_s}\xspace}

\def\SlowPi  		{\ensuremath{\pion_s}\xspace}

\newcommand{\sqseq}[1]{\ensuremath{\sqs =\text{#1}}\xspace}

\def\c          {\ensuremath{c}\xspace}

\newcommand{\magnitude}[1]{\ensuremath{\left|{#1}\right|}\xspace}
\newcommand{\magsq}[1]{\ensuremath{\magnitude{#1}^2}\xspace}

\newcommand{\tene}[1]{\ensuremath{10^{#1}}\xspace}
\newcommand{\xtene}[2]{\ensuremath{#1 \times \tene{#2}}\xspace}

\def\CPV{\CP-violation\xspace}
\def\C	{\ensuremath{C}\xspace}

\def\CPV     {\CP violation\xspace}

\def\pp      {\ensuremath{\proton\proton}\xspace}
\def\dzhphm{\decay{\Dz}{\hphm}}

\title{Indirect $\boldsymbol{\CP}$ Violation in $\boldsymbol{\decay{\Dz}{\hphm}}$ Decays at \lhcb}

\ShortTitle{Measurement of \agamma at \lhcb}

\author{\speaker{Michael Alexander}\thanks{On behalf of the \lhcb collaboration.}\\
  University of Glasgow\\
  E-mail: \email{michael.alexander@glasgow.ac.uk}}

\abstract{Indirect \CPV in the \Dz system can be probed by measuring the parameter \agamma, defined as the \CP asymmetry of the effective lifetime of the \Dz meson decaying to a \CP eigenstate. This can be significantly enhanced beyond Standard Model predictions by new physics. Measurements of \agamma using \dzkk and \dzpipi decays reconstructed from \pp collisions collected by the \lhcb experiment, corresponding to an integrated luminosity of 1.0 \invfb, are presented. The results are
\begin{align*}
      \agamma(\pion\pion) &= \xtene{(+0.33 \pm 1.06 \pm 0.14)}{-3}, \nonumber \\
      \agamma(\kk) &= \xtene{(-0.35 \pm 0.62 \pm 0.12)}{-3}, \nonumber 
\end{align*}
where the uncertainties are statistical and systematic, respectively. These are the most precise measurements of their kind to date, and show no evidence of \CPV.}

\FullConference{The 15th International Conference on B-Physics at Frontier Machines at the University of Edinburgh,\\
		14 -18 July, 2014\\
		University of Edinburgh, UK}

\begin{document}

\section{Introduction}

\begin{sloppypar}
Similarly to the \Bz and \Bs systems, the mass eigenstates of the \Dz system, \ket{\PD_{1,2}}, with masses $m_{1,2}$ and widths $\Gamma_{1,2}$, are superpositions of the flavour eigenstates \mbox{$\ket{\PD_{1,2}} = p \ket{\Dz} \pm q \ket{\Dzb}$}, where $p$ and $q$ are complex and satisfy \mbox{$\magsq{p} + \magsq{q} = 1$}. This causes mixing between the \ket{\Dz} and \ket{\Dzb} states, and allows for ``indirect'' \CPV in mixing, and in interference between mixing and decay, when decaying to a \CP eigenstate. Indirect \CP asymmetries in the \Dz system can be significantly enhanced beyond Standard Model (SM) predictions by new physics \cite{Bobrowski_indirectCPVCharm2010}. In decays of \Dz mesons to a \CP eigenstate $f$, indirect \CPV can be probed using \cite{aGammaYCPTheory}
\begin{equation*}
\agamma \equiv \agammadefnot \approx \agammaexp,
\end{equation*}
where \gammahat is the inverse of the effective lifetime of the decay, \etacp is the \CP eigenvalue of $f$, \mbox{$x \equiv 2(m_2 - m_1)/(\Gamma_1 + \Gamma_2)$}, \mbox{$y \equiv (\Gamma_2 - \Gamma_1)/(\Gamma_1 + \Gamma_2)$}, \mbox{$A_m\equiv(|q/p|^2-|p/q|^2)/(|q/p|^2+|p/q|^2)$}, \mbox{$A_d\equiv(|A_f|^2-|\bar{A}_{f}|^2)/(|A_f|^2+|\bar{A}_{f}|^2)$}, with $\optionalBar{A}_{f}$ the decay amplitude, and \mbox{$\phi \equiv arg(q\bar{A}_{f}/p A_f)$}. The effective lifetime is defined as the average decay time of a particle with an initial state of \ket{\Dz} or \ket{\Dzb}, \ie that obtained by fitting the decay-time distribution of signal with a single exponential.
\end{sloppypar}

The \lhcb detector at the \lhc, \cern, is a forward-arm spectrometer, specifically designed for high precision measurements of decays of $b$ and \c hadrons \cite{JINST_LHCb}. During 2011 the experiment collected \pp collisions at \sqseq{7 \tev} corresponding to an integrated luminosity of 1.0 \invfb. Due to the large \ccbar production cross section \cite{lhcb_promptCharmProduction2013}, the decay-time resolution of approximately 50 \fs for \Dz decays \cite{lhcb_veloPerformance2014} and the excellent separation of $\pi$ and \kaon achieved by the detector \cite{lhcb_richPerformance2014}, it is very well suited to measure \agamma with high precision.

\section{Methodology}

\begin{figure}
  \centering
  \includegraphics[width=0.35\textwidth]{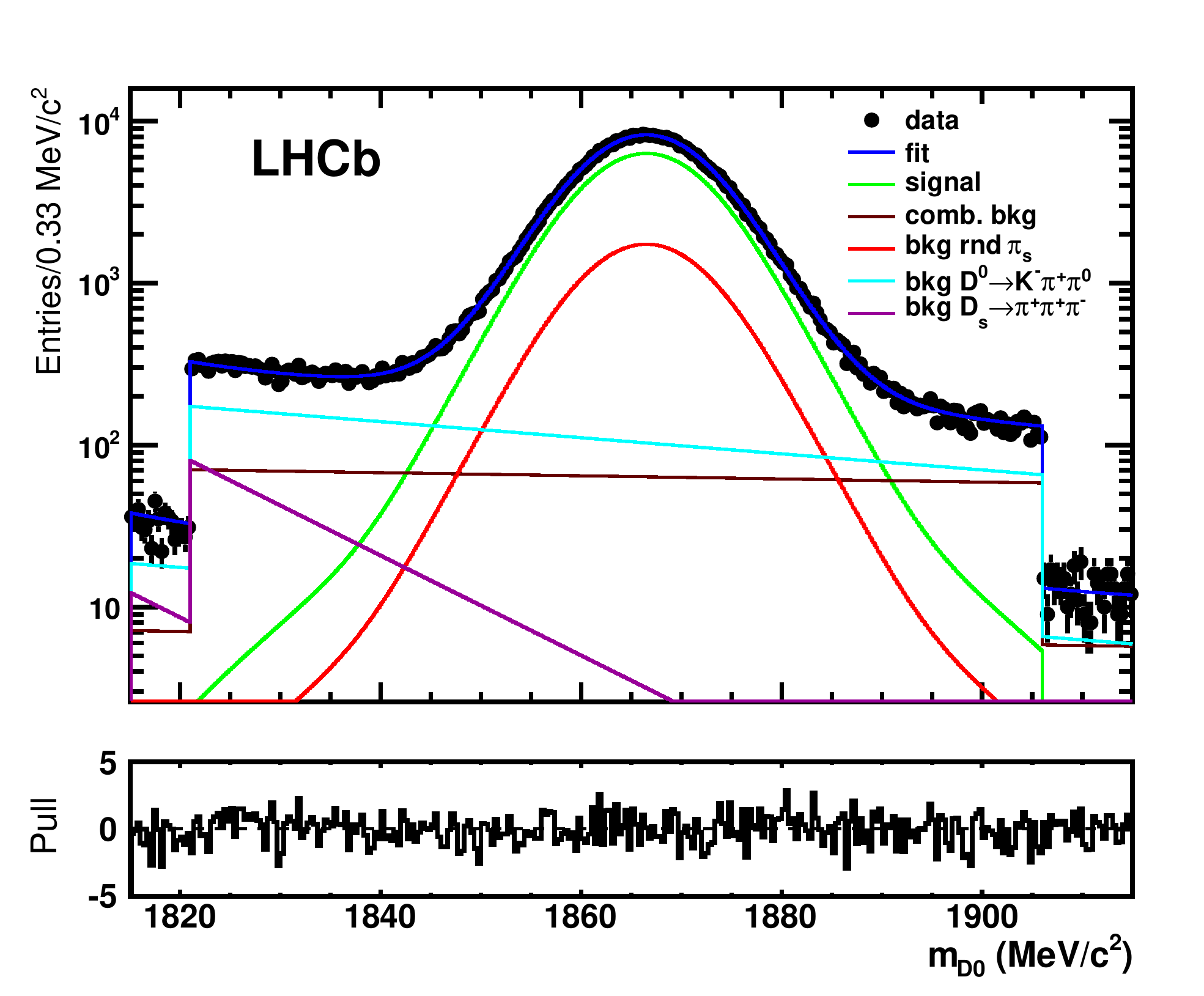} \qquad
  \includegraphics[width=0.35\textwidth]{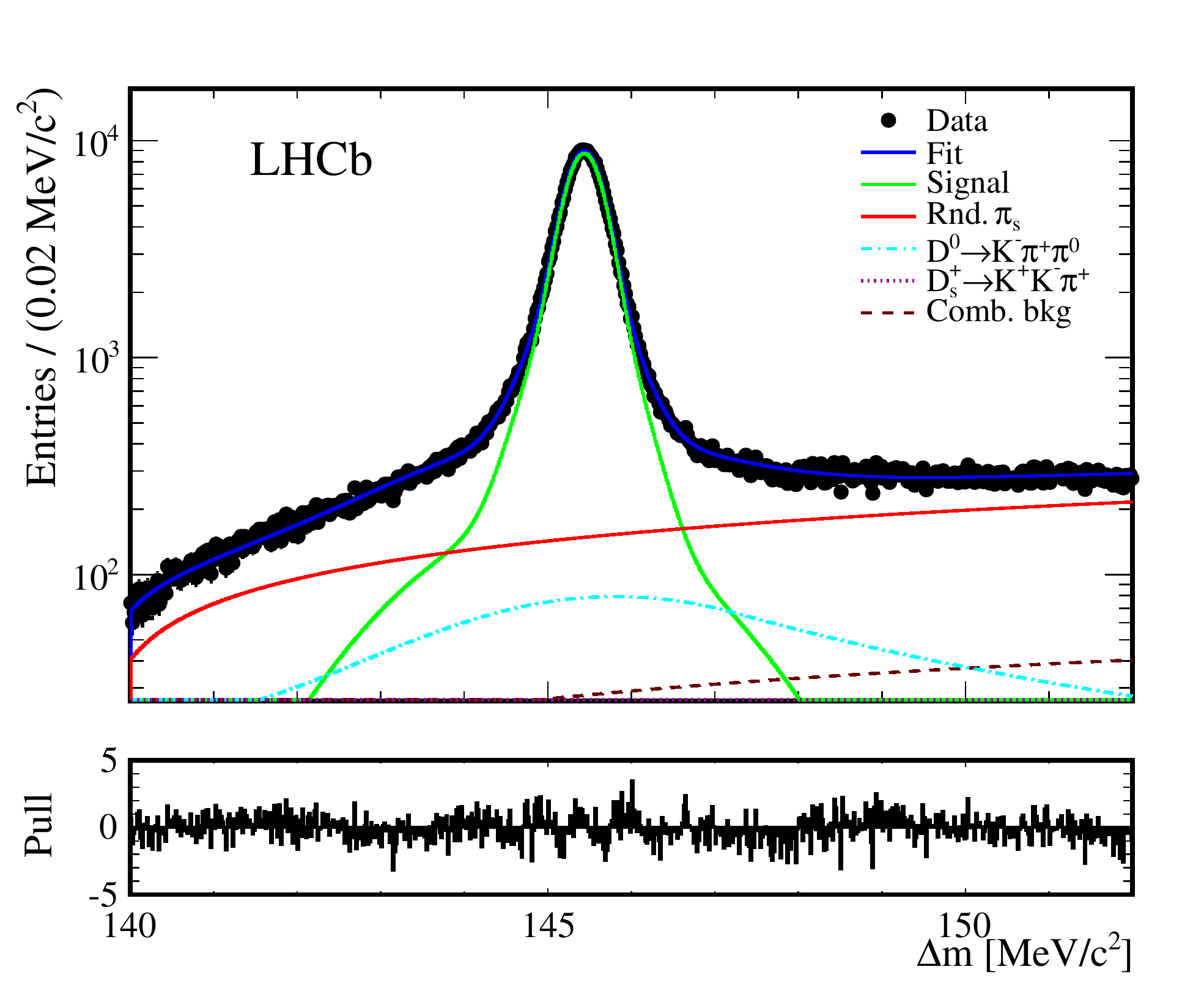}
  \caption{Fits to (left) the \Dz invariant mass distribution and (right) the \mbox{$\deltam\equiv m(\Dstarp) - m(\Dz)$} distribution for \dzkk candidates from the data subset with magnet polarity down, recorded in the earlier of the two running periods.}
  \label{fig:massfits}
\end{figure}

The decay chain \dstdzpi is used to determine the flavour of the \Dz candidates at production, via the charge of the \SlowPi meson. The \CP-even \KpKm and \pipi final states are used to calculate \agamma \cite{lhcb_agamma2014}. The predominant candidate selection criteria require the \KpKm or \pipi tracks to have large impact parameter (IP), large transverse momentum (\pt), invariant mass within 50 \mev of the world average \Dz mass, and for the vector sum of their momenta to point closely back to the position of the \pp collision. Using data corresponding to an integrated luminosity of 1.0 \invfb, 4.8M \dzkk candidates and 1.5M \dzpipi candidates are selected. The data are divided by \Dz flavour, the polarity of the \lhcb dipole magnet, and two separate running periods. Combinatorial and partially reconstructed backgrounds are discriminated using a simultaneous fit to the distributions of \Dz mass and \mbox{$\deltam\equiv m(\Dstarp) - m(\Dz)$}. Examples of these fits are shown in Fig. \ref{fig:massfits} for \dzkk candidates, for data recorded with the magnet polarity down during the earlier of the two running periods.

\begin{figure}
  \centering
  \includegraphics[width=0.35\textwidth]{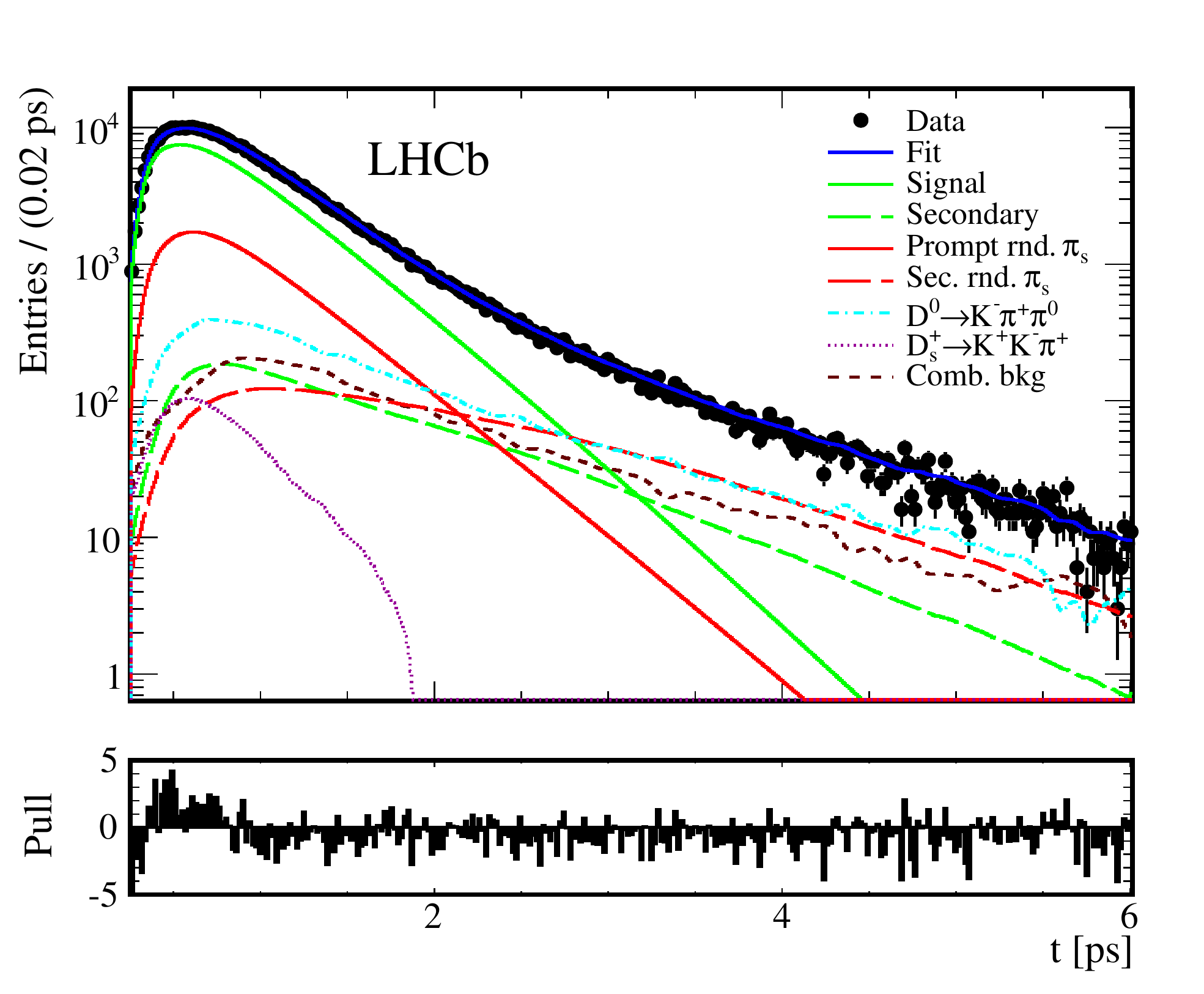} \qquad
  \includegraphics[width=0.35\textwidth]{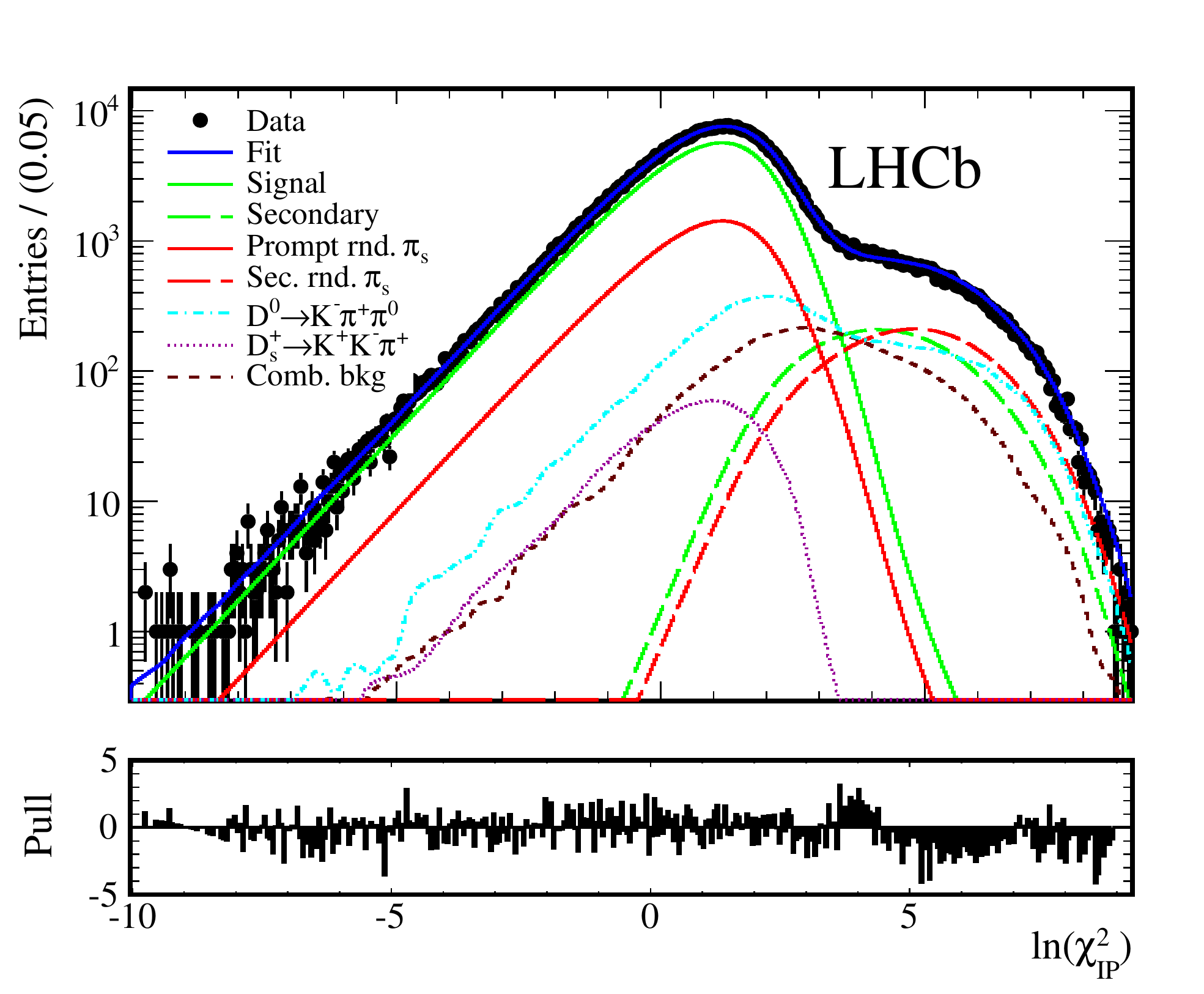}
  \caption{Fits to (left) the \Dz decay-time distribution and (right) the $\ln(\chisq_{\text{IP}})$ distribution for \dzkk candidates from the data subset with magnet polarity down, recorded in the earlier of the two running periods.}
  \label{fig:timefits}
\end{figure}

A fit to the decay-time distribution of the candidates is then used to determine the effective lifetimes of the \Dz and \Dzb signal. Only candidates for which the \Dstarp is produced directly at the \pp collision are considered as signal. The background from \decay{\PB}{\Dstarp \text{X}} decays is discriminated by simultaneously fitting the distributions of the decay time and the natural logarithm of the \chisq of the hypothesis that the \Dz candidate originates directly from the \pp collision ($\ln(\chisq_{\text{IP}})$). The selection efficiency as a function of decay time is obtained from data using per-candidate acceptance functions, as described in detail in Ref. \cite{lhcb_yCPAGamma}. The decay-time and $\ln(\chisq_{\text{IP}})$ distributions for combinatorial and specific backgrounds are obtained from the data using the discrimination provided by the mass and \deltam fits to employ the $_{\text{s}}$Weights technique \cite{sPlots} with kernel density estimation \cite{scott_densityEstimation}. Figure \ref{fig:timefits} shows fits to the distributions of decay time and $\ln(\chisq_{\text{IP}})$ for \dzkk candidates, using the same data subset as Fig. \ref{fig:massfits}. Inaccuracies in the fit model are examined as a source of systematic uncertainty, as discussed in the following section.

\section{Results and systematics}

\begin{figure}
  \centering
  \includegraphics[height=0.18\textheight]{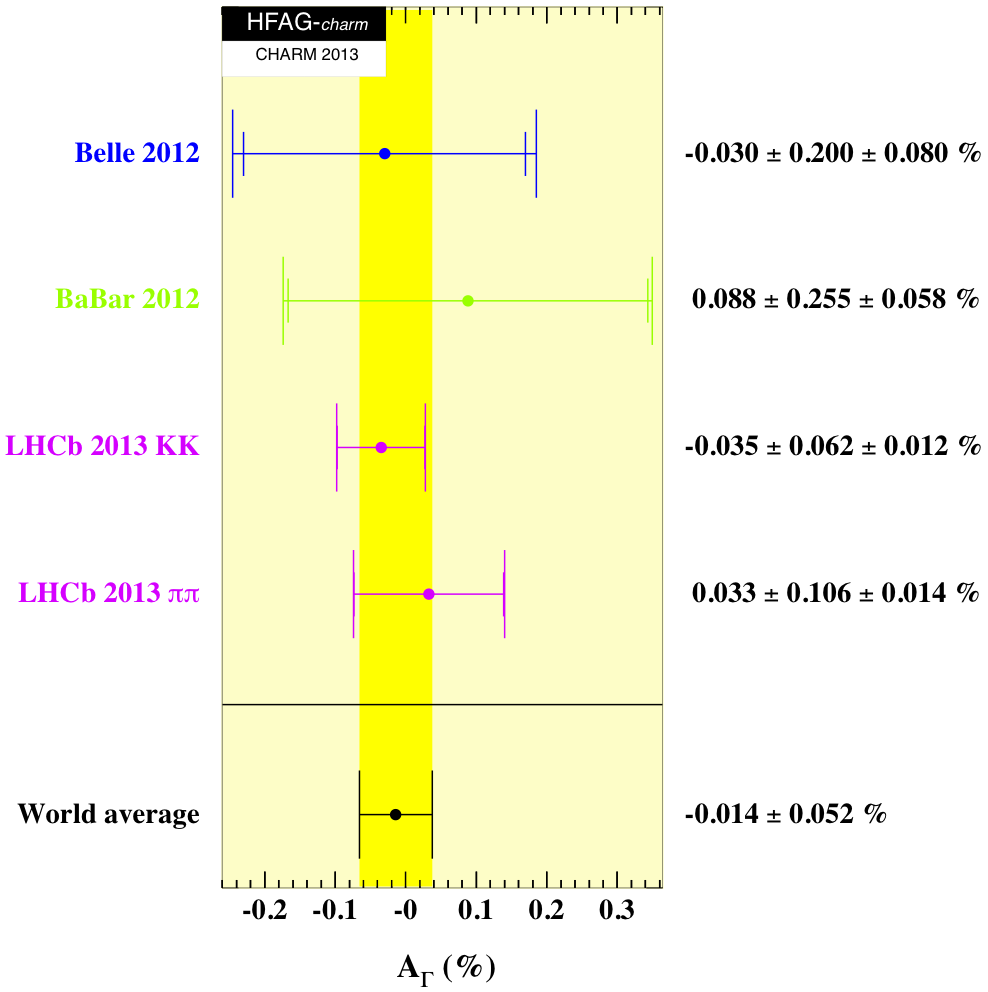}\qquad
  \includegraphics[height=0.18\textheight]{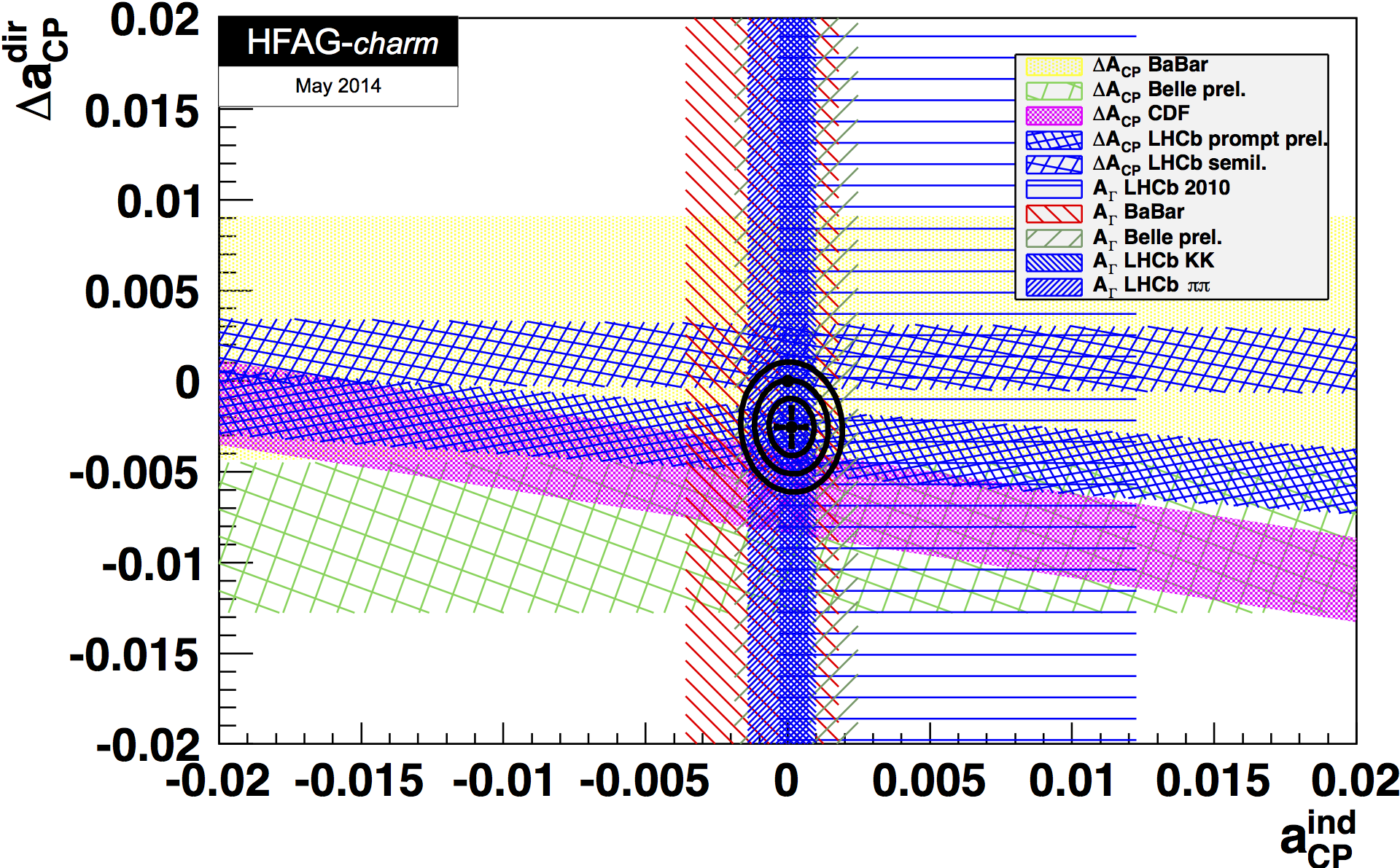}
  \caption{The world averages of (left) \agamma and (right) direct vs. indirect \CPV in \dzhphm decays, reproduced from \cite{HFAG2014}.}
  \label{fig:averages}
\end{figure}

The fits detailed in the previous section find
\begin{align*}
      \agamma(\pion\pion) &= \xtene{(+0.33 \pm 1.06 \pm 0.14)}{-3}, \nonumber \\
      \agamma(\kk) &= \xtene{(-0.35 \pm 0.62 \pm 0.12)}{-3}, \nonumber
\end{align*}
where the uncertainties are statistical and systematic, respectively. These are the most precise measurements of their kind to date, and show no evidence of \CPV. The dominant systematic uncertainties arise from the modelling of the selection efficiency as a function of decay time, and the modelling of the background from \decay{\PB}{\Dstarp \text{X}} decays. Figure \ref{fig:averages} (left) shows the world average of \agamma, which is dominated by these measurements and is consistent with zero. Figure \ref{fig:averages} (right) shows the combined fit to measurements of direct and indirect \CPV in \dzhphm decays, which yields a p-value for zero \CPV of 5.1 \% \cite{HFAG2014}. 

The precision of these measurements will be improved by the addition of 2.1 \invfb of data already collected during 2012. Together with data to be recorded in run II, and, in time, following the \lhcb upgrade, measurements with precisions of approximately \xtene{1}{-4} are possible, giving great potential for the discovery of indirect \CPV in the \Dz system.

\bibliographystyle{unsrt}
\bibliography{bibliography}

\begin{thebibliography}{10}

\bibitem{Bobrowski_indirectCPVCharm2010}
M.~Bobrowski, A.~Lenz, J.~Riedl, and J.~Rohrwild.
\newblock {How Large Can the SM Contribution to CP Violation in $D^0-\bar D^0$
  Mixing Be?}
\newblock {\em JHEP}, 1003:009, 2010.

\bibitem{aGammaYCPTheory}
{M. Gersabeck, M. Alexander, S. Borghi, V. V. Gligorov, C. Parkes}.
\newblock {On the interplay of direct and indirect CP violation in the charm
  sector}.
\newblock {\em J. Phys. G: Nucl. Part. Phys.}, 39, 2012.
\newblock 045005.

\bibitem{JINST_LHCb}
The~LHCb collaboration.
\newblock {The LHCb Detector at the LHC.}
\newblock {\em JINST}, 3(S08005), 2008.

\bibitem{lhcb_promptCharmProduction2013}
The~LHCb collaboration.
\newblock Prompt charm production in pp collisions at.
\newblock {\em Nuclear Physics B}, 871(1):1 -- 20, 2013.

\bibitem{lhcb_veloPerformance2014}
The LHCb~VELO group.
\newblock {Performance of the LHCb Vertex Locator}.
\newblock {\em JINST}, 9:09007, 2014.

\bibitem{lhcb_richPerformance2014}
The LHCb~RICH group.
\newblock Performance of the lhcb rich detector at the lhc.
\newblock {\em The European Physical Journal C}, 73(5), 2013.

\bibitem{lhcb_agamma2014}
The~LHCb collaboration.
\newblock {Measurements of indirect CP asymmetries in $D^0\to K^-K^+$ and
  $D^0\to\pi^-\pi^+$ decays}.
\newblock {\em Phys.Rev.Lett.}, 112(4):041801, 2014.

\bibitem{lhcb_yCPAGamma}
The~LHCb collaboration.
\newblock {Measurement of mixing and CP violation parameters in two-body charm
  decays}.
\newblock {\em JHEP}, 1204:129, 2012.

\bibitem{sPlots}
{Muriel Pivk and Francois R. Le Diberder}.
\newblock {sPlot: a statistical tool to unfold data distributions}.
\newblock {\em Nucl. Instrum. Meth. A}, 555:356--369, 2005.

\bibitem{scott_densityEstimation}
David~W. Scott.
\newblock {\em {Multivariate Density Estimation: Theory, Practice, and
  Visualization}}.
\newblock Wiley Series in Probability and Mathematical Statistics. {John Wiley
  \& Sons, Inc.}, 1992.

\bibitem{HFAG2014}
{The Heavy Flavor Averaging Group.}
\newblock \url{http://www.slac.stanford.edu/xorg/hfag/}, August 2014.

\end{thebibliography}

\end{document}